# Quantum Diamond Microscopy for Non-Destructive Failure Analysis of an Integrated Fan-Out Package-on-Package iPhone Chip

Bartu Bisgin[1], Marwa Garsi[1], Andreas Welscher[1], Michael Hanke[1], Fleming Bruckmaier[1]

[1]QuantumDiamonds GmbH, Friedenstraße 18, 81671 Munich, Germany

## Abstract

The increasing complexity of advanced semiconductor packages, driven by chiplet architectures and 2.5D/3D integration, challenges conventional failure localization methods such as lock-in thermography (LIT) and complicates current Failure Analysis (FA) workflows. Dense redistribution layers and buried interconnects limit the ability of established techniques to understand failure mechanisms non-destructively. In this work, we validate quantum diamond microscopy (QDM) based on nitrogen-vacancy (NV) centers in diamond as a non-destructive localization method through magnetic current path imaging at the package level. Using commercial Integrated Fan-Out Package-on-Package (InFO-PoP) devices from iPhones, we showcase a complete FA workflow that includes QDM to localize a short-type failure at an Integrated Passive Device (IPD) at the package backside. We showcase that the QDM results provide invaluable information on top of conventional techniques and can significantly enhance root-cause identification in package-level FA workflows. This work demonstrates the potential of QDM for broader integration into semiconductor chip and package analysis workflows.

## Introduction

As the semiconductor industry evolves toward higher-density interconnects and heterogeneous integration, package-level fault localization becomes increasingly challenging [1], [2]. Modern semiconductor packages often combine multiple active dies, memory stacks, and complex interconnects within extremely compact geometries. These architectures introduce multiply-routed current paths, buried redistribution layers, and three-dimensional heat dissipation profiles that obscure traditional optical and thermal fault signatures [3].

Techniques such as lock-in thermography (LIT), photo-emission microscopy (PEM), Thermally-Induced Voltage Alteration (TIVA), and Optical-Beam Induced Resistance Change (OBIRCH) have long been the first line of investigation in failure analysis [4], [5], [6] [7], [8]. In advanced packages, signals from multiple layers can overlap or interfere, making it difficult to attribute a hotspot to a specific interconnect, redistribution trace, or die-to-die interface [1]. The growing structural complexity of 2.5D and 3D architectures calls for novel techniques that can go beyond working with heat and emissions, enabling novel ways of looking into the electrical activity within buried layers [9].

Quantum diamond microscopy (QDM) is a promising technique that allows magnetic current imaging (MCI) of electronics, utilizing Nitrogen-Vacancy (NV) centers [10] in diamond [11], [12], [13]. An investigation of the technique for a certain type of short-failure has been reported in [14]. The technique combines wide-field optics with a diamond quantum sensor, allowing for non-destructive, quantitative current-path imaging over large chip and package areas. Unlike scanning methods such as magnetic force microscopy (MFM) [15] or superconducting quantum interference devices (SQUIDs) [16], the QDM can read 3D magnetic field maps at high speed, large field of view, with high resolution, and high sensitivity, while operating at ambient conditions (room temperature and pressure). With these features combined, the QDM enables the precise extraction of lateral and depth information from imaged electrical activity, making it industrially relevant for semiconductor analysis [17].

In this work, we demonstrate a seamless integration of the QDM technique into the failure analysis workflow of commercial, advanced integrated fan-out package-on-package (InFO-PoP) devices, with the clear value proposition of giving exact current paths leading up to a failure within an IPD, allowing the failure analysis engineers to correlate novel data to their physical analysis.

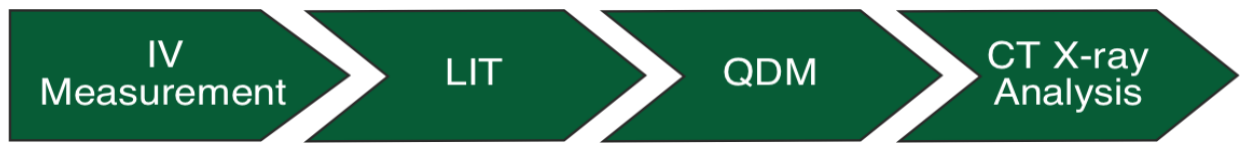

**Figure 1:** The failure analysis workflow employed in this work.

Using two packages, one functioning as intended and one defective, we correlate and compare the novel magnetic data obtained by the QDM with a LIT hotspot, as well as virtual cross-sections through Computed Tomography X-Ray (CT) on the failing package. The results highlight QDM's ability to distinguish conductive pathways that are invisible to traditional methods and point toward a Power-Ground short within the device due to a short failure. Overall, the framework employed in this work demonstrates a new workflow that matches active failure current paths to the layout and physical analysis, unlocking novel insights. This study establishes QDM as a powerful, non-destructive diagnostic method for analyzing next-generation semiconductor packages, unlocking novel insights for fault isolation and root cause analysis.

## Experimental Setup

The experimental setting employed a realistic package-level failure analysis workflow combining electrical testing with LIT, followed by the non-destructive QDM, as well as a CT-scan for destructive physical analysis. The workflow can be found in Figure 1. An illustration of the QDM is shown in Figure 2 (a). Two devices were measured: a so-called "Known Good Die" (KGD) and the failing part. Backside analysis has been given focus. This setup allowed illustrating the value that QDM brings to real-world FA workflows, and has shown the complementarity between conventional methods and the magnetic imaging capabilities introduced by the QDM.

### Device Under Test (DUT)

The device under test was an Apple A12 package, which integrates the system-on-chip (SoC), the graphics processing unit (GPU), and the neural processing unit (NPU) within a single package. The A12 is manufactured using the Integrated Fan-Out Package-on-Package (InFO-PoP) architecture offered by TSMC [18]. A representative package cross-section is shown in Figure 2 (b). InFO-PoP is a wafer-level fan-out packaging technology that allows direct integration of logic and memory dies through high-density redistribution layers (RDLs) and through integrated vias (TIVs). By eliminating the need for an organic substrate and C4 bumps, the package achieves a thinner profile, reduced electrical parasitics, and improved thermal performance compared to traditional flip-chip PoP designs. This architecture allows tight coupling between the application processor and dynamic random-access memory (DRAM), enhancing signal integrity and power efficiency for mobile devices [19].

In this work, two A12 packages were extracted from commercial iPhone components and prepared for the failure analysis experiments. The devices were carefully desoldered from their native printed circuit boards and re-wired to allow independent biasing of the SoC and DRAM power connections. For this purpose, a custom rewiring board was used with a rectangular opening at its center. Device 1 was simply soldered as is with remaining balls, while Device 2 had to be back polished. Each package was fixed within this opening using an adhesive layer, providing mechanical stability while leaving both sides of the device accessible. This configuration enabled wire bonding from the backside while still allowing access from the front for subsequent LIT and QDM measurements on both sides. An illustration and a picture of the rewiring configuration can be found in Figure 2 (c).

Although the A12 originates from an earlier iPhone generation (2018), its InFO-PoP structure remains representative of current technology, as the same architecture is still widely employed in modern devices [20], [21]. Its adoption is driven by the efficient electrical and thermal performance of InFO-PoP, as well as its suitability for thin, high-performance mobile designs. The results presented here are therefore directly relevant to today's state-of-the-art implementations.

### Electrical Testing

First, electrical tests were performed to verify the biasing conditions of the A12 packages. Defined voltage sweeps were applied through the re-wiring interface while monitoring the total current response, obtaining the I-V curves (see Supplementary Materials).

These initial measurements ensured stable operation with the rewiring configuration and reproducible electrical conditions for subsequent optical and magnetic characterization.

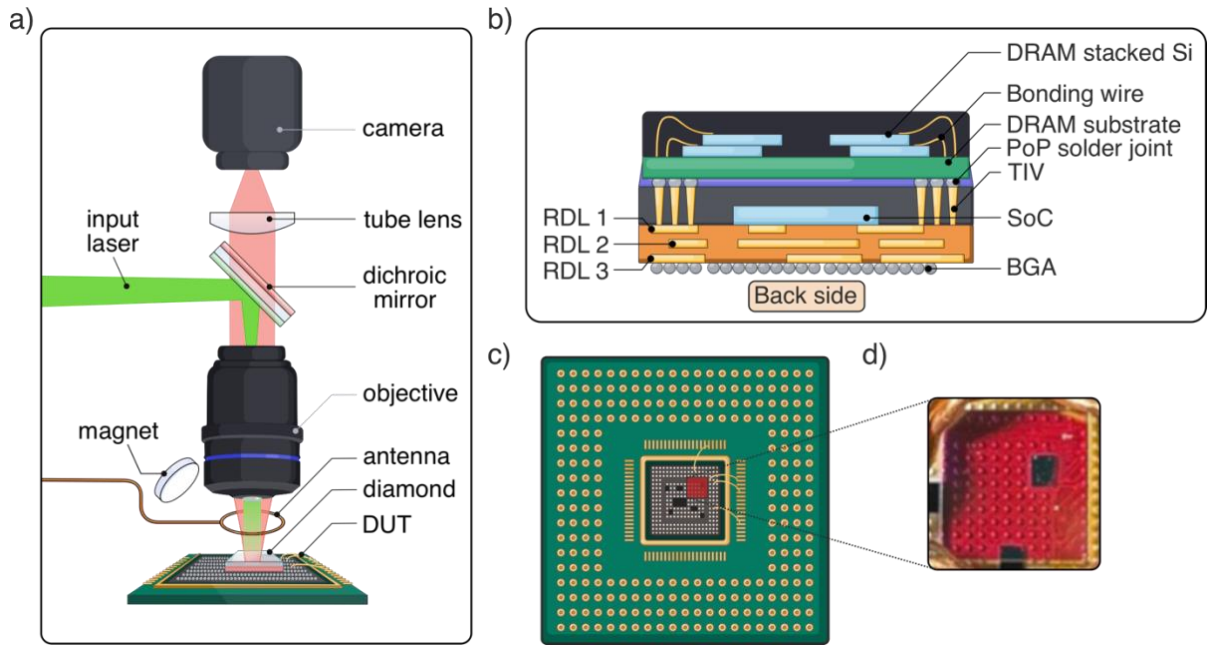

**Figure 2: a)** QDM device diagram. A 532 nm laser is reflected by a dichroic mirror through an objective onto the NVs in the diamond plate on the DUT (NV layer in red). A permanent magnet lifts NV degeneracy, while a microwave antenna drives transitions. Red fluorescence is collected through the objective, filtered, and imaged by the camera. **b)** Illustration of an InFO-PoP cross-section. Bottom to top, Ball Grid Array (BGA), Redistribution Layers (RDL) that connect the balls to the internals, System-on-Chip (SoC), Through InFO Vias (TIV), Bumps that connect the top and bottom packages (PoP solder joint), Dynamic Random Access Memory (DRAM) substrate, and dies with bond wires. **c)** Illustration of the DUT, with the InFO-PoP sample re-wired to a PCB. The red rectangle illustrates the placement of the diamond. **d)** The inset shows a real image from the measurement. Only the sensor is visible to not reveal sensitive information about the package.

## Lock-in Thermography (LIT)

Lock-in thermography was employed as a conventional technique for package-level fault localization. The measurements were performed on the ELITE tool from ThermoFisher. The re-wired devices were mounted under an infrared microscope and electrically biased using the same conditions later applied during the QDM measurements. A periodic voltage modulation was applied to generate localized Joule heating in defective regions, and the resulting thermal emission was recorded synchronously with the excitation signal. By demodulating the infrared response at the driving frequency, amplitude and phase images were obtained, highlighting areas of excess power dissipation within the package. The resulting hotspots were used to define the regions of interest for subsequent QDM measurements, providing a non-destructive thermal reference for comparison.

## Quantum Diamond Microscopy (QDM)

All QDM measurements were performed on a prototype of the commercial quantum diamond microscope, the QDm.1, by QuantumDiamonds. Each measurement was performed by loading the A12 DUT into the instrument and positioning a 4 mm x 4 mm-sized diamond quantum sensor on top of a region of interest (RoI), with the NV side facing the DUT. A high-density ensemble of NVs was used, and the NV layer was illuminated with a 532 nm laser through the microscope objective in epifluorescence configuration, shown in Figure 2 (a).

When subjected to an external magnetic field, the energy levels of the NV center's $m_s = \pm 1$ states split due to the Zeeman effect. The energy (and hence, frequency) separation of the $m_s = \pm 1$ transitions from $m_s = 0$ is directly proportional to the magnetic-field projection along the NV orientation axes [10], [11]. Utilizing an ensemble, all four potential directions of NVs were present, and three were used to span the full space and construct the 3D magnetic maps.

We performed optically detected magnetic resonance (ODMR) by sweeping the microwave frequency and recording the corresponding changes in NV fluorescence. A decrease in fluorescence intensity indicates a resonance condition, allowing the local magnetic field to be quantified.

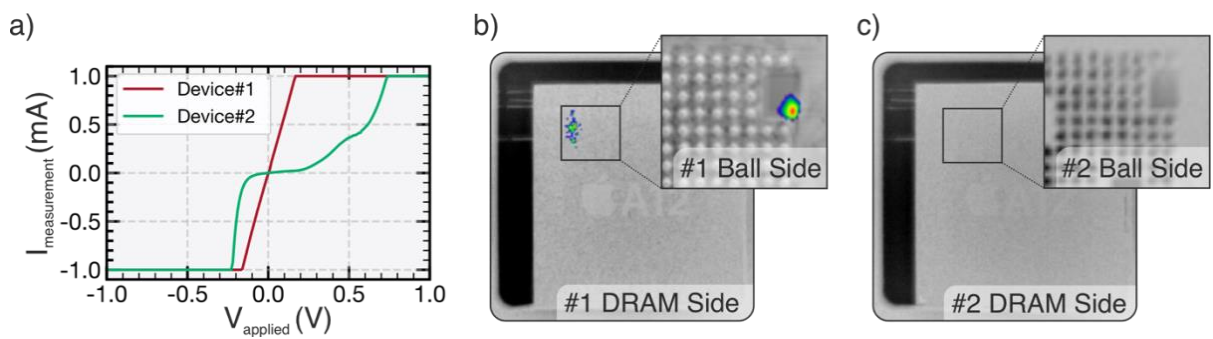

**Figure 3: a)** IV curves obtained from Device 1 (red) and Device 2 (green), with Device 1 showing a linear behavior consistent with short failure. **b-c)** Front-side and backside LIT images of Device 1 and 2, respectively. Device 1 shows a hotspot on both sides, with the backside being much more localized, while Device 2 shows no signal.

The fluorescence intensity was imaged across the full field of view using a scientific camera, yielding an ODMR spectrum for every pixel. Each spectrum was analyzed to extract the resonance frequencies of all three NV orientations, and the magnetic field vector was reconstructed from their projections. It should be noted that, due to using ensembles, every pixel of the camera contains a multitude of NV centers in all four orientations, which allows for inherent averaging within each pixel. A static bias field within the mT range was applied to lift the degeneracy between different orientation NV centers.

Magnetic field maps were acquired both with and without a constant current applied to the package to remove background contributions. The resulting differential maps reveal current-induced magnetic fields from the redistribution layers and interconnects within the package, forming the basis for quantitative current-path identification, which can aid in layout-informed physical analysis. A corresponding current density distribution can be reconstructed from the measured magnetic fields using the inverse Biot-Savart law, which relates the local magnetic field to the current density within the conducting layers [22], [23]. This allows quantitative mapping of current flow in buried interconnects with micrometer-scale precision.

## Computed Tomography X-Ray

Computed Tomography X-Ray (CT) imaging was performed as the final step in the analysis workflow to obtain structural information on the package for physical failure analysis. The measurements were carried out using a Versa XRM520 system by ZEISS operated in high-resolution mode. The package was carefully mounted in the instrument, and a detailed three-dimensional image of the complete device was acquired.

Virtual cross-sections were generated along the *x*- and *y*-axes through the IPD region of interest to inspect the internal structure in the area identified by the preceding analyses. As the technique involves exposure to X-ray radiation, it is categorized as a destructive physical analysis method. The CT imaging provided detailed geometrical insight into the internal layers and interfaces of the package to support subsequent root-cause investigation.

## Results

Electrical characterization was first carried out to establish the behavior of both A12 packages under defined bias conditions. The I-V response of Device 1, shown in Figure 3 (a), exhibited a linear characteristic, indicative of a resistive short within the package. In contrast, Device 2 showed the expected nonlinear curve associated with nominal device operation. This initial measurement suggested that Device 1 contained a low-resistance conduction path at the package level, motivating subsequent localization experiments.

To spatially identify the origin of this behavior, both devices were examined using LIT. Under the same bias conditions, Device 1 revealed a distinct thermal hotspot, visible from both the front-side and back-side of the package, and most pronounced from the back (see Figure 3 (b)). On the backside, the hotspot location has been observed at the corner of one of the IPDs. Device 2, by comparison, exhibited no measurable thermal contrast within the detection sensitivity of 10 µK (see Figure 3 (c)). The location of the backside hotspot in Device 1 was therefore defined as the RoI for both devices, for further magnetic imaging using the QDM.

a) Device #1: Defective Package

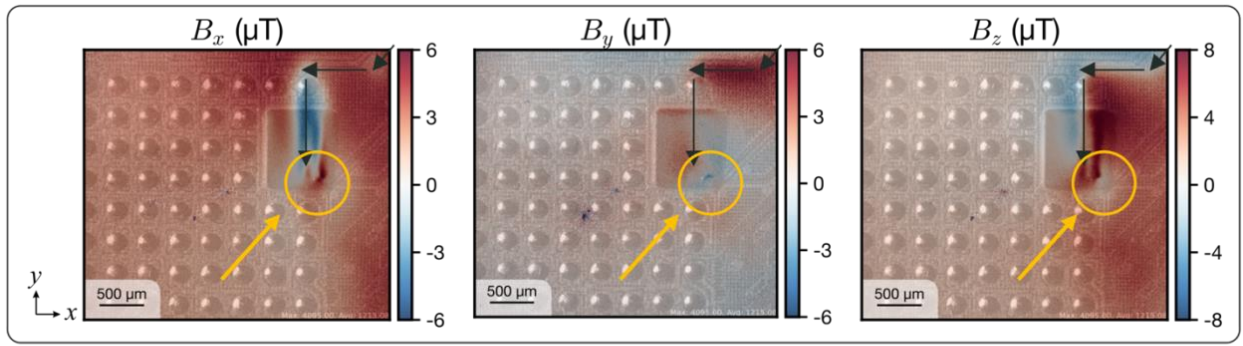

b) Device #2: Known-Good-Die

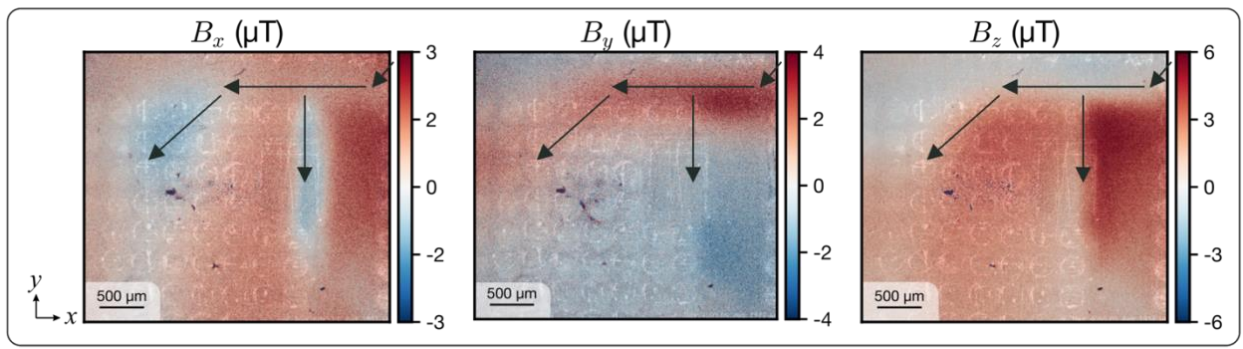

**Figure 4:** QDM measurement output. All 3 images in each row are taken in a single measurement and separated into components in post-processing. Black arrows indicate the approximate current path, with $B_x$ highlighting activity mainly in the y-direction (blue for current travelling in -y), $B_y$ highlighting activity in the x-direction (red for current travelling in -x). $B_z$ is indicating the general xy-current distribution. **a)** Field components of the defective package are shown with an overlay on the optical image of the backside, taken through the diamond. A kink in $B_z$ is visible, indicated by the orange circle, which matches the spot found by LIT certifying the anomaly. **b)** Field components for "Known-

Good-Die". A different current path is inferred, branching to the IPD similar to a) and towards the left of the FoV, indicated approximately by black arrows. Contrary to a), there are no signs of the anomaly. Overall, the field strength is also lower compared to the defective package due to the distribution of current into a second branch.

With the diamond sensor positioned over the backside IPD RoI (refer to Figure 2 (d)), QDM measurements were performed on both packages under an identical bias of 100 µA. The measured magnetic field maps displayed a clear difference in the current flow patterns.

In Device 1, a strong and well-defined magnetic feature was observed (see Figure 4 (a)), most readily visible in the z component of the magnetic field $B_z$, corresponding to a single conductive trace extending toward and terminating at the hotspot identified by LIT. The magnitude and spatial gradient of the signal indicated a concentrated current path consistent with a short, low-resistance connection in one of the redistribution layers leading up to the IPD.

In Device 2, the same region produced a considerably weaker magnetic response (see Figure 4(b)). The observed magnetic map revealed that the same trace identified in Device 1 branched into an additional pathway for Device 2, resulting in a more diffuse magnetic pattern. The branching in Device 2 suggests that the nominal current divides between multiple conductive paths in the KGD, whereas in the defective device, the current is redirected through a single, unintended low-resistance path.

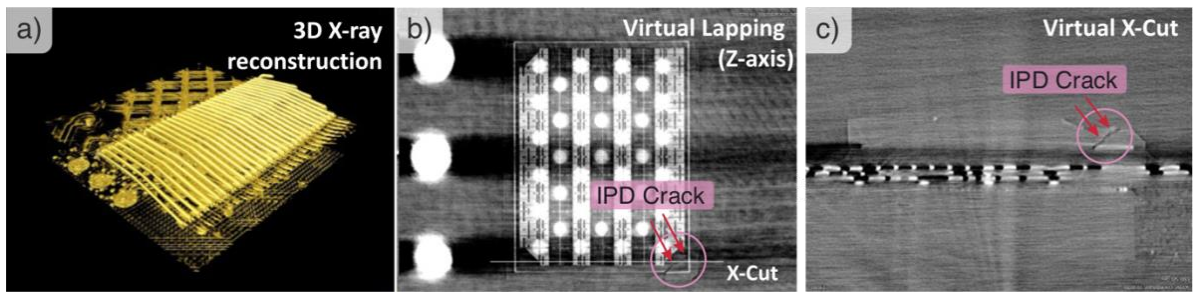

**Figure 5:** 3D CT X-ray images. **a)** Showing a large area of the package with multiple levels and with wire bonds visible at the device side. The BGA are also visible at the bottom, with smaller balls above showcasing the level which corresponds to the IPD. **b)** Virtual cross-section along z showing inside the IPD with distinct power and ground rails, matching up to current path findings of QDM. A crack is visible on the lower right corner, and an X-cut is taken. **c)** x-cut reveals the 3D extent of the vertical crack, suggesting a physical power-to-ground short between rails.

Computed tomography cross-sections of the failing device revealed a clear physical defect at the corner of the IPD (see Figure 5). The reconstructed images showed local chipping and a vertical crack extending through the IPD, intersecting metallized layers associated with the power and ground lines. This structural damage indicates a direct short between the two nets within the IPD. The location of the crack coincides precisely with the region where QDM detected the kink and termination of the current path, confirming that the observed magnetic anomaly originates from a physical fracture acting as the failure site. In addition, the QDM magnetic maps clearly resolved the individual power and ground lines within the region, distinguishing the conductive paths converging at the hotspot- an insight unattainable through thermal emission alone.

Taken together, the electrical, thermal, magnetic, and structural analyses form a consistent picture of the underlying defect mechanism. The linear I-V behavior of Device 1 indicated the presence of an internal short; LIT localized the associated dissipative hotspot, and QDM directly visualized the buried current path terminating at this location. Subsequent CT imaging confirmed a physical crack within the IPD at the same site, providing structural evidence of a vertical short between power and ground connections. The complementary reference measurement on Device 2 verified that the observed magnetic and thermal contrasts originate from altered electrical connectivity rather than from geometrical or measurement artifacts. These combined results demonstrate that quantum diamond microscopy can non-destructively reveal buried current paths and pinpoint failure locations with a level of diagnostic detail that, when correlated with physical analysis, surpasses the capabilities of conventional thermal fault-localization techniques.

## Conclusion and Outlook

This study demonstrates the application of quantum diamond microscopy as a novel, non-destructive tool for failure localization in advanced semiconductor devices. By combining electrical testing, LIT, QDM, and computed tomography on commercial A12 InFO-PoP devices, a comprehensive analysis of the failure was obtained. While electrical measurements indicated a low-resistance path and thermal imaging localized an associated hotspot, QDM uniquely visualized the underlying current distribution within buried layers for both the non-defective and defective samples. The ability to image samples regardless of defectivity is a unique property of magnetic imaging, and has potential other use-cases such as side-channel attacks [24], [25].

The magnetic maps revealed distinct current behavior between the two devices, providing direct spatial insight into conductive pathways that remain inaccessible to conventional optical or thermal methods. Subsequent computed tomography imaging confirmed the presence of a crack within the IPD, resulting in an internal short between the power and ground connections. These results highlight how QDM can non-destructively identify current paths and add novel insights on top of the hotspot, which can be used for layout-informed physical analysis.

The proposed workflow with Electrical Testing, LIT, QDM, and CT demonstrates a powerful approach for advanced failure analysis. Expanding this approach to additional device types and different biasing schemes will further clarify the versatility of QDM in mapping current paths and help localize defects under realistic failure analysis conditions. Extending the framework to package types such as HBM stacks, 2.5D interposers, and chiplet assemblies could help establish QDM as a standard technique for non-destructive semiconductor FA on top of generic hotspot analysis. Especially when depth information is needed, QDM could uniquely identify which layer a certain electrical defect is located in. Detailed investigation of such a depth localization case can be the subject of further work.

Furthermore, because QDM acquires magnetic information from all pixels simultaneously, the same instrument can be extended beyond static imaging to capture device dynamics as time-resolved magnetic maps. With appropriate AC biasing or pulsed excitation, this approach can enable open-failure detection and transient current tracking, as well as MHz-GHz field measurements using NV-based AC magnetometry. Such capabilities would unlock new applications ranging from dynamic failure analysis to side-channel analysis and hardware-security investigations in integrated circuits.

## Acknowledgements

We are grateful to the Integrated Service Technology (iST) failure analysis team for their support in the realistic test-case preparation, electrical characterization, lock-in thermography, and CT X-ray measurements reported in this work.

## Bibliography

[1] Electronic Device Failure Analysis Society, Ed., *Electronic Device Failure Analysis Technology Roadmap*. ASM International, 2023. doi: 10.31399/asm.tb.edfatr.9781627084628.
[2] IEEE International Roadmap for Devices and Systems, "Metrology," 2023, *Institute of Electrical and Electronics Engineers*. doi: 10.60627/FF6X-D213.
[3] L. C. Wagner, "Failure analysis challenges," in *Proceedings of the 2001 8th International Symposium on the Physical and Failure Analysis of Integrated Circuits. IPFA 2001 (Cat. No.01TH8548)*, Singapore: IEEE, 2001, pp. 36–41. doi: 10.1109/IPFA.2001.941451.

[4] O. Breitenstein, W. Warta, and M. C. Schubert, *Lock-in Thermography: Basics and Use for Evaluating Electronic Devices and Materials*, vol. 10. in Springer Series in Advanced Microelectronics, vol. 10. Cham: Springer International Publishing, 2018. doi: 10.1007/978-3-319-99825-1.

[5] O. Breitenstein, "Thermal Failure Analysis by IR Lock-in Thermography," *Microelectron. Fail. Anal. Desk Ref. Sixth Ed.*, vol. Sixth Edition, pp. 330–339.

[6] O. Breitenstein, J. P. Rakotoniaina, F. Altmann, J. Schulz, and G. Linse, "Fault Localization and Functional Testing of ICs by Lock-in Thermography," presented at the ISTFA 2002, Phoenix, Arizona, USA, Oct. 2002, pp. 29–36. doi: 10.31399/asm.cp.istfa2002p0029.

[7] E. I. Cole, J. M. Soden, J. L. Rife, D. L. Barton, and C. L. Henderson, "Novel failure analysis techniques using photon probing with a scanning optical microscope," in *Proceedings of 1994 IEEE International Reliability Physics Symposium RELPHY-94*, San Jose, CA, USA: IEEE, 1994, pp. 388–398. doi: 10.1109/RELPHY.1994.307808.

[8] E. I. Cole, P. Tangyunyong, D. A. Benson, and D. L. Barton, "TIVA and SEI developments for enhanced front and backside interconnection failure analysis," *Microelectron. Reliab.*, vol. 39, no. 6–7, pp. 991–996, June 1999, doi: 10.1016/S0026-2714(99)00136-5.

[9] "International Roadmap for Failure Analysis (IRFA) - Solutions to Gaps Summary." ASM International. [Online]. Available: https://www.asminternational.org/istfa-2025/technical/international-roadmap-for-failure-analysis/?srsltid=AfmBOorh9iNU377NoZHEec1_5zbGu2EerB7k9fOVTPSTXzLpAhrwn12s

[10] M. W. Doherty, N. B. Manson, P. Delaney, F. Jelezko, J. Wrachtrup, and L. C. L. Hollenberg, "The nitrogen-vacancy colour centre in diamond," *Phys. Rep.*, vol. 528, no. 1, pp. 1–45, July 2013, doi: 10.1016/j.physrep.2013.02.001.

[11] E. V. Levine *et al.*, "Principles and techniques of the quantum diamond microscope," *Nanophotonics*, vol. 8, no. 11, pp. 1945–1973, Nov. 2019, doi: 10.1515/nanoph-2019-0209.

[12] M. Garsi *et al.*, "Three-dimensional imaging of integrated-circuit activity using quantum defects in diamond," *Phys. Rev. Appl.*, vol. 21, no. 1, p. 014055, Jan. 2024, doi: 10.1103/PhysRevApplied.21.014055.

[13] A. Nowodzinski, M. Chipaux, L. Toraille, V. Jacques, J.-F. Roch, and T. Debuisschert, "Nitrogen-Vacancy centers in diamond for current imaging at the redistributive layer level of Integrated Circuits," *Microelectron. Reliab.*, vol. 55, no. 9–10, pp. 1549–1553, Aug. 2015, doi: 10.1016/j.microrel.2015.06.069.

[14] P. Kehayias, J. Walraven, A. L. Rodarte, and A. M. Mounce, "High-Resolution Short-Circuit Fault Localization in a Multilayer Integrated Circuit Using a Quantum Diamond Microscope," *Phys. Rev. Appl.*, vol. 20, no. 1, p. 014036, July 2023, doi: 10.1103/PhysRevApplied.20.014036.

[15] A. Pu, A. Rahman, D. Thomson, and G. E. Bridges, "Magnetic force microscopy measurement of current on integrated circuits," in *IEEE CCECE2002. Canadian Conference on Electrical and Computer Engineering. Conference Proceedings (Cat. No.02CH37373)*, Winnipeg, Man., Canada: IEEE, 2002, pp. 439–444. doi: 10.1109/CCECE.2002.1015265.

[16] Y. Kaneko and A. Orozco, "Current Imaging using SQUID and GMR Sensors for Failure Analysis," *Proc. 37th Annu. NANO Test. Symp. NANOTS2017*.

[17] M. Garsi, A. Welscher, B. Bisgin, and M. Hanke, "Quantum Diamond Microscopy for Semiconductor Failure Analysis", [Online]. Available: https://static.asminternational.org/edfa/202502/18/

[18] TSMC, "Integrated Fan-Out (InFO) Wafer Level Packaging." [Online]. Available: https://3dfabric.tsmc.com/english/dedicatedFoundry/technology/InFO.htm

[19] TSMC, "TSMC 3DFabric® for Mobile." [Online]. Available: https://www.tsmc.com/english/dedicatedFoundry/technology/platform_smartphone_tech_WLSI

[20] ChipWise, "Apple A18 & A18 pro die shot." [Online]. Available: https://chipwise.tech/our-portfolio/apple-a18-a18-pro-die-shot/

[21] TechNode Feed, "iPhone 16e to feature A18 chip with TSMC's 3nm process and custom 5G chip." [Online]. Available: https://technode.com/2025/02/22/iphone-16e-to-feature-a18-chip-with-tsmcs-3nm-process-and-custom-5g-chip/

[22] B. J. Roth, N. G. Sepulveda, and J. P. Wikswo, "Using a magnetometer to image a two-dimensional current distribution," *J. Appl. Phys.*, vol. 65, no. 1, pp. 361–372, Jan. 1989, doi: 10.1063/1.342549.

[23] J.-P. Tetienne, N. Dontschuk, D. A. Broadway, A. Stacey, D. A. Simpson, and L. C. L. Hollenberg, "Quantum imaging of current flow in graphene," *Sci. Adv.*, vol. 3, no. 4, p. e1602429, Apr. 2017, doi: 10.1126/sciadv.1602429.

[24] Y. Zhou and D. Feng, "Side-Channel Attacks: Ten Years After Its Publication and the Impacts on Cryptographic Module Security Testing," *IACR Cryptol. ePrint Arch.*, vol. 2005, [Online]. Available: https://api.semanticscholar.org/CorpusID:9365379

[25] J. N. Lenz, S. K. Perryman, D. J. Martynowych, D. A. Hopper, and S. M. Oliver, "Hardware Trojan Detection Potential and Limits with the Quantum Diamond Microscope," *ACM J. Emerg. Technol. Comput. Syst.*, vol. 21, no. 1, pp. 1–24, Jan. 2025, doi: 10.1145/3711712.